\documentclass{JHEP}

\usepackage{amssymb,amsfonts,amsmath,amscd,latexsym,epsfig}
\input{epsf}

\newcommand {\be} {\begin{equation}}
\newcommand {\ee} {\nonumber \end{equation}}
 \newcommand {\ber}{\begin{eqnarray*}}
 \newcommand {\eer} {\end{eqnarray*}}
\newcommand {\bea}{\begin{eqnarray}}
 \newcommand {\eea} {\end{eqnarray}}

\def \p {\partial}

\def \R {\mathbb{R}}

\def \o {\omega_{ij}}
\def \hg {\hat{g}_{ij}}
\def \p {\partial}
\def \xm {X^\mu}
\def \L  {L_{BI}}
\def \ephi {e^{-\Phi}}
\def \e {\epsilon}
\def \d {\delta}
\def \half {\frac{1}{2}}
\def \vf {\varphi}
\def \KK {\p_\alpha K \p^\alpha K }
\def \Pdd {P^{\alpha\beta}\p_\alpha\vf\p_\beta\vf}

\title{D3-branes in NS5-brane backgrounds}
\input{epsf}
\usepackage{epsfig}
\author{ Sylvain Ribault
\\Centre de Physique Th{\'e}orique \footnote{Unite mixt{\'e} du CNRS et de l'
Ecole Polytechnique, UMR7644}
\\Ecole Polytechnique, 91128 Palaiseau, France
\\ E-mail: \email{ribault@cpht.polytechnique.fr}
} \abstract{ We study D3-branes in an NS5-branes background
defined by an arbitrary 4d harmonic function. Using a gauge-invariant
formulation of Born-Infeld dynamics as well as the supersymmetry condition, we
find the general solution for the $\omega$-field. We propose
an interpretation in terms of the Myers effect.
}

\preprint{ 
 CPHT RR 003.0103}

\begin{document}

\section{Introduction and summary}

Any configuration of parallel NS5-branes creates a non-trivial string
background, described by the following fields on the transverse four
dimensions~:
\ber
G_{\mu\nu}= V \delta_{\mu\nu} \\
H_{\mu\nu\rho}=\e_{\mu\nu\rho\sigma}\p_\sigma V \\
e^{2\Phi}=V
\eer
Here, $V$ can be any harmonic function of the four transverse
coordinates $X^\mu$. 

These backgrounds play an important r{\^o}le in (little) string theory;
they are
related to various exact string backgrounds. For instance, the
near-horizon geometry of $k$ superposed NS5-branes is $\R_\Phi\times
SU(2)$, where $\R_\Phi$ is the linear dilaton background. The
near-horizon geometry of NS5-branes spread on a circle \cite{sfetsos} 
is related by T-duality
to an orbifold of $SL(2,\R)/U(1)\times SU(2)/U(1)$. 
In \cite{tong}, an NS5-branes background is used to
exhibit the effect of worldsheet instantons on T-duality.
NS5-branes spread on a three-sphere provide another
interesting configuration, with a dilaton everywhere finite
\cite{kkpr}. 

All this motivates the study of D-brane probes in such
backgrounds. Such probes have already been used in some particular cases
of NS5-branes background \cite{egkrs,pelc} and in some U-dual
configurations \cite{savv,pauli}.
First, the D1-branes (we mean branes extending along one of the four
transverse dimensions, with an unspecified number of flat directions
 parallel to the
NS5-branes) are not affected by
the NS5-branes and are straight lines \footnote{This can be seen from their
Born-Infeld action: the $V$ factors coming from the dilaton and
metric cancel each other.}. On the contrary, the D3-branes can take
quite complicated shapes \cite{pelc}, 
in relation with the Hanany-Witten effect \cite{hana}. 

In this note we investigate the general properties of D3-branes in \emph{all}
such backgrounds.
Let us summarize the results. First,
an useful tool to study the shapes of D3-branes will be the
gauge-invariant 
rewriting
of the Born-Infeld equations of motion, Eq. (\ref{BIinv}). We will
briefly comment on the geometrical significance of this rewriting in
terms of a non-symmetric second fundamental form. Then we will write
the Born-Infeld and SUSY equations for D3-branes in general NS5-branes
backgrounds. Those two equations turn out to be equivalent. 
We will find the general solution for
the $\omega$-field on the brane, Eq. (\ref{omegasol}). This $\omega$-field
describes the D1-brane charge of the D3-brane, enabling us to
speculate about the D3-branes being formed as bound
states of D1-branes via a kind of Myers effect \cite{myers}. 

\section{Invariant Born-Infeld equations of motion}

The Born-Infeld action reads
\ber
S_{BI}(\xm,F)=\int dx^i \L =\int dx^i \ephi\sqrt{\det{(\hg+\o)}},
\eer
where $\o=\hat{B}_{ij}+F_{ij}$ is the gauge-invariant worldvolume
two-form, subject to the constraint $d\omega=\hat{H}$ where $dB=H$.
 The action is gauge-invariant, as well as the equation of motion for
 the $F$-field $E^k=-\p_i
\frac{\d\L}{\d F_{ik}}$; 
but not the equation of motion for
 the embedding $\xm(x^i)$, that is $E_\mu=\frac{\d\L}{\d
   \xm}-\p_i\frac{\d\L}{\d \p_i\xm}$. However, it is possible to
add a combination of $E ^k$ to the equation $E_\mu$ and to obtain an
equivalent, gauge-invariant equation. This was already done in
\cite{sken}, where the equation 
$
E^\mu-E^jB_\nu{}^\mu\p_jX^\nu=0
$
was used
\footnote{
An other way
  of deriving 
this equation is to
introduce a contravariant worldvolume three-form $C^{ijk}$ in order to
impose $d\omega=\hat{H}$, and to consider the action
$S'(\xm,\omega,C)=\int dx^i \ephi\sqrt{\det{(\hg+\o)}}+\lambda\int dx^i
C^{ijk}(d\omega-\hat{H})_{ijk}$.
It is then possible to eliminate $C^{ijk}$ from the equations of
motion of $\xm$, using the equations of $\omega$, now considered as a
dynamical field.  
}. 
Here we propose a different combination, which will turn out to
have a much more interesting geometrical interpretation~:
\bea
E^\mu+E^j(\omega_j{}^k\p_kX^\mu-B_\nu{}^\mu\p_jX^\nu)=0.
\eea
Indeed this equation may be rewritten
\bea
-\sqrt{\det{(\hat{g}+\omega)}}
[(\hat{g}+\omega)^{-1}]^{ji}\left(\p_i\p_j\xm +\Gamma
  ^\mu_{\nu\rho}\p_iX^\nu \p_jX^\rho-\hat{\Gamma}^k_{ij}\p_k\xm\right)
\nonumber \\
-\sqrt{\det{(\hat{g}+\omega)}}\left(\p^\mu\Phi-\hat{g}^{ij}\p_i\Phi\p_jX^\mu
\right) =0,
 \label{BIinv} \eea
where we used the spacetime connection
\bea
\Gamma^\mu_{\nu\rho}=\Gamma (g)^\mu_{\nu\rho}-\half
H^\mu{}_{\nu\rho},
\eea
and the induced worldvolume connection
\bea
\Gamma ^k_{ij}=\Gamma(\hat{g})^k_{ij}-\half \hat{H}^k{}_{ij}.
\eea
The equation (\ref{BIinv}) involves the following two-form with values
in the tangent space
\bea
\Omega ^\mu_{ij}=\p_i\p_j\xm +\Gamma
  ^\mu_{\nu\rho}\p_iX^\nu \p_jX^\rho-\hat{\Gamma}^k_{ij}\p_k\xm.
\eea
This generalizes the second fundamental form and shares its basic
properties. Indeed our $\Omega$ is transverse ($\Omega
^\mu_{ij}\p_kX^\mu =0$), and it satisfies generalized Gauss-Codazzi
equations 
\bea
R(\hat{\Gamma})_{ijkl}=R(\Gamma)_{ijkl}+g_{\mu\nu}\Omega
^\mu_{[lj}\Omega ^\nu_{k]i} \\
(R_N)_{ij}{}^{\mu\nu}=R(\Gamma)^{\mu\nu}{}_{ij}-\hat{g}^{kl}\Omega_{jk}^{[\mu}
\Omega_{il}^{\nu]}
\eea
where $R_N$ is the curvature of the spin connection
$\omega_i^{ab}=\half\xi_\mu ^{[a}(\p_i+\Gamma^\mu_{i\nu})\xi^{\nu b]}$
for some orthonormal basis $\xi^{\mu a}$ of the normal space;
explicitly we have
\ber
 (R_N)_{ij}{}^{\mu\nu}=\left(\p_{[i}\omega_{j]}^{ab}- \omega
  _{[i}^{ac}\omega _{j]}^{bc}\right)\xi^\mu_a\xi^\nu_b.
\eer
Thus, we were able to reformulate the Born-Infeld equations of
motion in a gauge invariant manner, using the connection with torsion
$\Gamma$ and the associated second fundamental form on the brane. This
suggests that those objects should contribute to the derivative 
corrections to the
Born-Infeld action when the B-field is present, generalizing purely
gravitational terms of \cite{bbg,foto}. More generally, this points to the
relevance of the connection $\Gamma$ for the D-branes geometry, as was
already noted in \cite{hkk}. 

For the moment, we will only use the gauge invariance of
eq. (\ref{BIinv}) in order to study D3-branes in an NS5-branes
background, without having to fix a gauge for the B-field or to find
an F-field on the brane.

\section{The case of D3-branes}

In order to write the equations which determine the geometry of a D3-brane
and its worldvolume two-form $\o$, let us define this geometry by the
equation $K(\xm)=$cst\ and write
the
 most general
local solution to the $F$-field Born-Infeld equation of motion $E^k$:
\bea
\o=V\e_{ijk}\frac{\p^k\vf}{\sqrt{1-\hat{g}^{mn} 
\p_m\vf\p_n\vf}}.      
\eea
Here $\vf$ is some function on the brane,
and we normalize
$\e_{ijk}$ so that it is a tensor,
$\epsilon_{123}=\sqrt{\det{\hat{g}}}$, where 
we define $\hat{g}_{ij}=\p_iX^\mu\p_jX_\mu$. We raise spacetime
indices with $\delta_{\mu\nu}$, not with the metric
$G_{\mu\nu}=V\delta_{\mu\nu}$. Thus, the worldvolume metric
$\hat{g}_{ij}$, with which we raise indices, does not coincide
with the standard induced metric $V\p_iX^\mu\p_jX_\mu$.

The unknown functions $K$ and $\vf$ are subject to two equations,
which we write using the projector onto the brane
\ber
P_{\mu\nu}=\delta_{\mu\nu}-\frac{\p_\mu K\p_\nu K}{\p^\rho K\p_\rho K}.
\eer
First, we have the gauge-invariant Born-Infeld equation 
\bea
\left( P^{\mu\nu}+\frac{P^{\mu\mu'}\p_{\mu'}\vf
  P^{\nu\nu'}\p_{\nu'}\vf}{1-P^{\alpha\beta}\p_\alpha\vf\p_\beta\vf} \right)
&& \p_\mu\p_\nu K 
+ V^{-1}\p_\mu V \p_\mu K \nonumber \\
-&& \frac{\sqrt{\p_\alpha K \p^\alpha
    K}}{\sqrt{1-P^{\alpha\beta}\p_\alpha\vf\p_\beta\vf}}
P^{\mu\nu}V^{-1}\p_\mu V\p_\nu \vf =0. \label{BI}
\eea
Second, we should not forget the 
equation $d\o=\hat{H}$~: 
\bea
&&\left( P^{\mu\nu}+\frac{P^{\mu\mu'}\p_{\mu'}\vf
  P^{\nu\nu'}\p_{\nu'}\vf}{1-P^{\alpha\beta}\p_\alpha\vf\p_\beta\vf}
\right)
\p_\mu\p_\nu K 
-\frac{\p_\alpha K \p^\alpha K}{\p_\alpha K \p^\alpha \vf}
\left( P^{\mu\nu}+\frac{P^{\mu\mu'}\p_{\mu'}\vf
  P^{\nu\nu'}\p_{\nu'}\vf}{1-P^{\alpha\beta}\p_\alpha\vf\p_\beta\vf}
\right)
\p_\mu\p_\nu \vf \nonumber \\
&+&\frac{\sqrt{\p_\alpha K \p^\alpha
    K}\sqrt{1-P^{\alpha\beta}\p_\alpha\vf\p_\beta\vf} }{\p_\alpha
    K \p^\alpha \vf }  V^{-1}\p_\mu V \p_\mu K - 
\frac{\p_\alpha K \p^\alpha K}{\p_\alpha K \p^\alpha \vf} 
P^{\mu\nu}V^{-1}\p_\mu V\p_\nu \vf =0.  \label{cons}
\eea
Our two equations are second-order partial differential equations. 
It is possible to find a first order equation by studying the
supersymmetry condition for the brane, which will turn out to be
equivalent to the Born-Infeld equation. 
First,
the background preserves the following supersymmetries~:
\ber 
\xi=V^{\frac{1}{16}}\xi_0,\ \xi_0={\rm\ cst},\
\Gamma_{6789}\xi=-\xi_c.
\eer
Then the D3-brane SUSY condition is
\bea
-i\sqrt{\det{(1+\hat{g}^{-1}\omega)}} \xi =
\Gamma_{0\mu\nu\rho}\e^{ijk}\p_iX^\mu \p_jX^\nu \p_k X^\rho \xi +
\Gamma_{0\mu} \e^{ijk}\p_iX^\mu \omega_{jk} \xi_c.
\eea
This does not depend at all on the harmonic 
function $V$ defining the background. With our notations, this can be
rewritten in the form $-i\xi=v^\mu \Gamma_{0\mu}\Gamma_{6789} \xi$,
with 
\bea
v^\mu =\frac{\sqrt{1-\Pdd}}{\sqrt{\KK}}\p^\mu K+ P^{\mu \nu} \p_\nu \vf
\eea 
At any given point, the
existence of such a $\xi$ is guaranteed by the fact that
$v_\mu v^\mu =1$. However, one shoud not forget that $\xi$ should
always remain in the same direction, $\xi=V^{\frac{1}{16}}\xi_0$. Thus
the SUSY condition for our D3-brane amounts to 
\bea
v^\mu={\rm\ cst}. \label{susy}
\eea
We can use this equation to eliminate $\vf$ from our expressions. In
particular, the solution for the $\omega$-field is 
\bea
\o=V\frac{\sqrt{\KK}}{\p_\beta K v^\beta  } 
\e_{ijk}\p^kX_\mu v^\mu ,\ \ \ 
v^\mu\ {\rm cst.} \label{omegasol}
\eea
The equations (\ref{BI}) and (\ref{cons}) both take the form
\ber
\left(P^{\mu\nu}+\frac{\KK}{(v^\beta\p_\beta K)^2}P^{\mu\mu'}v_{\mu'
    }P^{\nu\nu'}v_{\nu'}\right)\p_\mu\p_\nu K + 
\left(\p^\mu K-\frac{\KK}{v^\beta\p_\beta K}P^{\mu\mu'}v_{\mu'}\right)
V^{-1}\p_\mu V =0.
\eer
Now that we have studied the local properties of the D3-branes, let us
say a word about the quantization conditions.
A quantized quantity is, as usual, defined for any two-cycle 
$S^2$ of the brane,
which is the boundary of some 3-surface $M$~:
\bea
I=\int_{M}H-\int_{S^2}\omega \label{RR}.
\eea
This quantity measures the RR charge of the D3-brane.

Three situations can happen~: first, if the background is created by
 localized individual NS5-branes, then the quantization is automatically
satisfied. Second, if the D3 passes through a stack of NS5, then the
angle at which the D3 emerges from this stack is quantized, like in
\cite{pelc}. Third, if the NS5 are
spread, then the quantization has to be added by hand. For instance
when the NS5 are spread on a circle, then the D3 has to intercept a
quantized portion of the circle.

\section{Examples and discussion}

In this last section we want to give a physical interpretation of our
results in terms of the Myers effect. Let us first mention a few
examples, where our equations can be solved or at least reduced to
differential equations.
\begin{itemize}
\item
Case when the background is asymptotically flat, $V\rightarrow 1$ at
infinity~: near infinity our D3-brane is nearly flat and 
we can solve the equations for its shape,
\bea
K=v_\mu X^\mu \left(V+\frac{\lambda}{r}+O(\frac{1}{r^3})\right),
\eea
with $V=1+\frac{kl_s^2}{r^2}+O(\frac{1}{r^3})$. The quantized
parameter $\lambda$ measures the number of D1s bound to the flat D3,
i.e. the D1-brane charge of the D3, as can be seen by computing the
quantity $I$, Eq. (\ref{RR}), near infinity. 
\item
$\R_\Phi\times SU(2)$: the background is defined by
$V=\frac{kl_s^2}{r^2}$, 
and the branes by
$K=\frac{v_\mu X^\mu}{r}$ for any constant $v_\mu$. 
These branes extend along the
whole linear dilaton direction, times a standard $S^2$ conjugacy class
in $SU(2)$ \cite{bds}.
\item Superposed NS5-branes: $V=1+\frac{kl_s^2}{r^2}$, the partial
  differential equation on $K$ reduces to a differential equation
  after assuming $K(v_\mu X^\mu,r)$ (for any constant $v_\mu$). 
See \cite{pelc} for more details.
\end{itemize}

\begin{figure}
\begin{center}
\epsfxsize=15cm \epsfbox{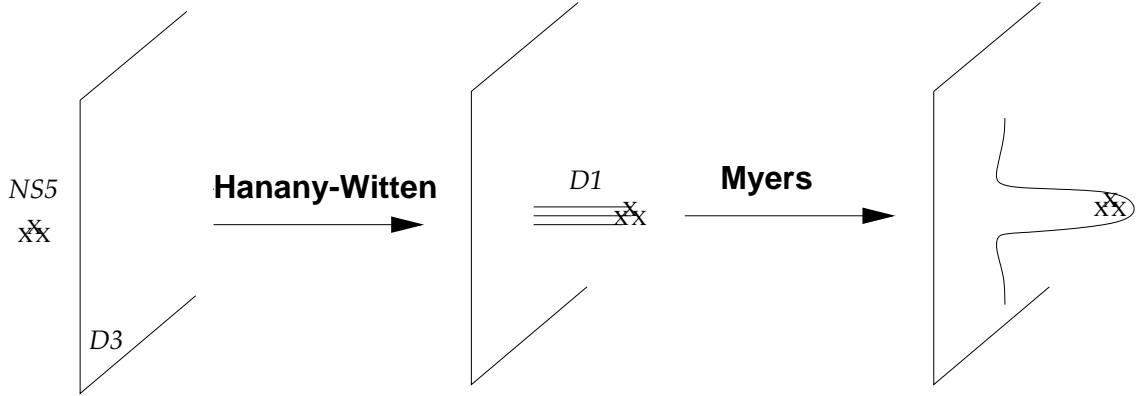}
\end{center}
\caption{A case of the proposed Myers effect} \label{fig}
\end{figure}

In the case of $\R_\Phi\times SU(2)$, the $S^2$ factor of
the brane can be formed from a stack of superposed D0-branes on $SU(2)$ 
 via the Myers effect. The $\R_\Phi\times SU(2)$ background
is the near-horizon geometry of the background defined by
superposed NS5-branes, and in this case the D3 should be considered as
a bound state of a flat D3 with a stack of D1-branes ending on the
NS5s (see Figure 
\ref{fig}). This is confirmed by computing its D1-brane
charge. However,
this only holds when the flat D3-brane does not go too far from the NS5s,
so that the throat of the corresponding curved D3 is not as thin as
the string length $\ell_s$. A limiting case occurs when we consider
D1-branes going to infinity without ending on any D3; then there is no
D3-brane solution which would be a candidate for a bound state of
those D1s, and the Myers effect does not occur. This might seem a bit
strange when we look at the near-horizon limit, but one should not
forget that this region suffers from a strong coupling problem
and cannot be expected to give reliable information. 
On the contrary, in the flat region far from the NS5s, the D1s
are not expected to form any bound state.

Now, we are led to conjecture that this phenomenon is general and that 
every supersymmetric D3-brane in an NS5-branes background
 is a bound state of D1s
(with or without a flat D3-brane at infinity depending on the background).
The main evidence we have is the existence of the constant vector
$v^\mu$, which indicates the direction of the D1-branes of
interest. Our D3 preserves the same supersymmetries as those D1-branes
and should have the correct charge, as hinted by Eq. (\ref{omegasol}).
It would be interesting to study this kind of Myers effect using
the nonabelian Born-Infeld action, however one should take into
account the fact that the original D1-branes may end on a D3-brane.

\acknowledgments{
I am grateful to Costas Bachas, Angelos Fotopoulos, 
Stefan Fredenhagen, Ruben Minasian and Marios Petropoulos
for interesting discussions and comments.
}


\renewcommand{\thesection}{A}
\setcounter{equation}{0}
\renewcommand{\theequation}{A.\arabic{equation}}
\appendix



\end{document}